\documentclass[runningheads]{llncs}
\usepackage{comment}
\usepackage{graphicx}
\usepackage{caption}

\begin{document}
\title{Towards a Predictive Patent Analytics and Evaluation Platform}
\author{Nebula Alam \and
Khoi-Nguyen Tran \and
Sue Ann Chen \and
John Wagner \and
Josh Andres \and
Mukesh Mohania
}
\authorrunning{Alam et al.}
\institute{IBM Research Australia, Southbank VIC 3006, Australia
\email{\{anebula,khndtran,sachen,john.wagner,josh.andres,mukeshm\}@au1.ibm.com}}
\maketitle              % typeset the header of the contribution
\begin{abstract}

The importance of patents is well recognised across many regions of the world. Many patent mining systems have been proposed, but with limited predictive capabilities. In this demo, we showcase how predictive algorithms leveraging the state-of-the-art machine learning and deep learning techniques can be used to improve understanding of patents for inventors, patent evaluators, and business analysts alike. Our demo video is available at \url{http://ibm.biz/ecml2019-demo-patent-analytics}


% Cross-Language Patent Mining, Patent Application, Patent Classification, Patent Information Retrieval, Patent Mining, Patent Valuation, Patent Visualization

\end{abstract}
%
%
% Sections are in their own files so we can work in parallel.

\section{Introduction}
\label{sec:intro}

Patents detail the innovations of individuals, organisations, and countries and represent the competitive landscape of ideas important to a society. Over the past decade, there has been a steady growth of intellectual property (IP) filing activity globally, averaging around 8$\%$ per year \cite{WIPO2018}. The main contributors of global growth in IP filings are China, USA, and Japan, contributing over 70$\%$ of patents out of 3.17 million that are filed worldwide in 2017 \cite{WIPOIPFacts2018}. By the end of 2018, more than 10 million patents have been issued by the United States Patent and Trademark Office (USPTO) \cite{USPTOFY2018}. The importance of patents is well recognised across many regions of the world and many patent mining systems have been proposed~\cite{Zhang2014}, but with limited predictive capabilities.

In this demo, we showcase how predictive algorithms using state-of-the-art machine learning and deep learning techniques can improve evaluation of patents for inventors, patent evaluators, and business analysts. %Our system is a work-in-progress that aims to provide novel ways of analysing patents using state-of-the-art machine learning techniques. We aim to improve the efficiency and effectiveness of patent evaluation by organising patents with respect to the technical features of the patents, and identifying sets of patents so as to fill capability gaps in existing products, and to create new business capabilities.
Currently, we have completed the engineering groundwork, summary statistics for inventors and for organisations, and a predictive algorithm of when filed patents will be issued by the patent office. This algorithm learns from the text of issued (i.e. granted) patents and estimates when published patents (i.e. in application stage) will be granted. In future demos and a technical publication, we will show how representational methods for patent text can significantly improve comparisons for search, modifying phrasing of ideas to shorten grant time, and identifying gaps in the knowledge coverage of patents and generating patent ideas from those gaps. Our patent data source is from the USPTO from 2005 to 2019 and a future technical publication will detail the data and algorithms used in this demo.

\subsection{Related Work}
\label{sec:relatedwork}

%We highlight a range of related work, from research into mining and classifying patents to deployed patent search and analysis systems.

%Mining and analysis of information from patent documents is a well-studied task which rapidly gained momentum due to the significant increase of patent data in the recent years.
%We list some of the existing patent analysis solutions that bear similarities to the system we propose here.

%%%%%%%%%%%%%%%%%%%%%%%%%%%%%%%%%%%%%%%%%%%%%%%%%

\textbf{Patent Systems.} There is a variety of existing proprietary and free online patent search systems with various capabilities.
%There are a variety of proprietary and free online patent search systems that currently exists with various capabilities.
%Free online patent search tools (e.g. Google Patent\footnote{\url{https://patents.google.com}}) undoubtly collect search information to improve their service and potentially to provide competitive advantage by analysing search terms from known IP address ranges.
Intellectual property offices (IPOs) around the world provide basic search capabilities on keywords, publication number, authors and date ranges\footnote{Espacenet: \url{https://www.epo.org/searching-for-patents/technical/espacenet.html}}\footnote{Patent Full-Text and Image Search: \url{https://www.uspto.gov/patent}}\footnote{PatentScope: \url{https://patentscope.wipo.int/search/en/search.jsf}}, as well as basic analytics tools and machine readable data.
%IP offices (IPOs) around the world provide basic search capabilities on keywords, publication number, authors, and date ranges.
%Some examples of search tools are Espacenet\footnote{\url{https://www.epo.org/searching-for-patents/technical/espacenet.html}} from the European Patent Office (EPO), Patent Full-Text and Image Search\footnote{\url{https://www.uspto.gov/trademarks-application-process/search-trademark-database}} from the United States Patent and Trademark Office (USPTO), and PatentScope\footnote{\url{https://patentscope.wipo.int/search/en/search.jsf}} from the World Intellectual Property Organization (WIPO). These IPOs also provide basic patent search and analytics tools, and machine readable data for companies use.
%AusPat\footnote{\url{http://pericles.ipaustralia.gov.au/ols/auspat/}} from IP Australia,
%Most IPOs provide access to machine readable patent data for free or for a fee.
Private patent search systems such as TotalPatent\footnote{\url{https://www.lexisnexis.com/totalpatent/}} build on this data to provide confidential capabilities (including search) for inventors and organisations. Public patent search systems such as lens.org\footnote{\url{https://www.lens.org/}} and Google Patents\footnote{\url{https://patents.google.com}} also make use of this data to provide additional commercial services or integration with other data sources to serve different market needs.

\textbf{Patent Mining.} A variety of patent mining techniques have been proposed as seen from a survey as reviewed by Zhang et al.~\cite{Zhang2014} and semantic methods based on text by Bonino et al. \cite{Bonino2010}. These techniques serve to extract information about technological innovations for businesses to make strategic investment \cite{Kyebambe2017} and to resolve problems inherent within the patent data for search and analysis.
%, such as complex natural language and ambiguous authors and citations.
Research based methods generally enrich the patent data, such as with metadata \cite{Agatonovic2008}, the classification taxonomy from the patent office \cite{Xiao2008}, derived data from the patent text and text summarisations \cite{Bruegmann2015}, informative data from patent citations and academic citations \cite{Julie2012}, patent domains \cite{Guo2009}, relationship with other patents \cite{Bergeaud2017}, and a combination of methods for forecasting technologies \cite{Kyebambe2017}.

\section{System Architecture}
\label{sec:system}

\begin{figure}[t]
\centering
\includegraphics[width=0.8\linewidth]{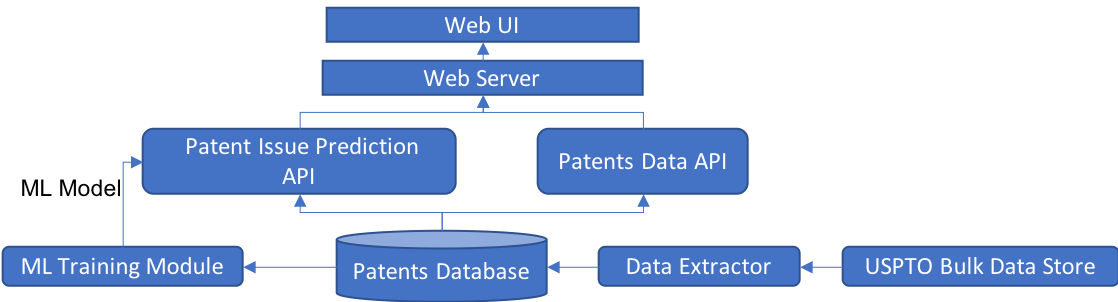}
\caption{System Architecture Overview}
\label{fig:system}
\end{figure}

 Figure \ref{fig:system} shows the high level architecture of the system. A \textbf{Data Extractor} module is responsible for extraction-transformation-loading (ETL) of issued and published patents from the \textbf{USPTO Bulk Data store}. This data is then processed to extract features to train models, where the best model is stored for deployment. The models are trained in the \textbf{ML Training Module} using \texttt{Scikit-learn}\footnote{\url{https://scikit-learn.org/stable/}} and Tensorflow\footnote{\url{https://github.com/tensorflow/tensorflow}} libraries. To create features from the patent text, we use a combination of manually engineered features, derived features, and representational features (e.g. \texttt{word2vec}\footnote{\url{https://www.tensorflow.org/tutorials/representation/word2vec}}, \texttt{fastText}\footnote{\url{https://fasttext.cc/}}). The prediction task in this demo is to estimate number of days until a patent will be issued, which includes a confidence score for each prediction made. A \textbf{Patent Issue Prediction API} is built around this model, exposing the predictive capability through a RESTful API, allowing end-users to estimate how soon their filed patents may be issued. Users can then examine other characteristics, such as the topic and domain of patents to determine factors affecting the estimated issued time. We provide a \textbf{Patent Data API} that encapsulates logic for querying the database, and provides end points for easy retrieval of selective and aggregated data, allowing direct access to the data for further analysis. A \textbf{Web User Interface} front-end (our demo) presents the analytics and prediction to the end-users.

\begin{figure}[t]
\centering
\begin{minipage}{.5\textwidth}
  \centering
  \includegraphics[width=.95\linewidth]{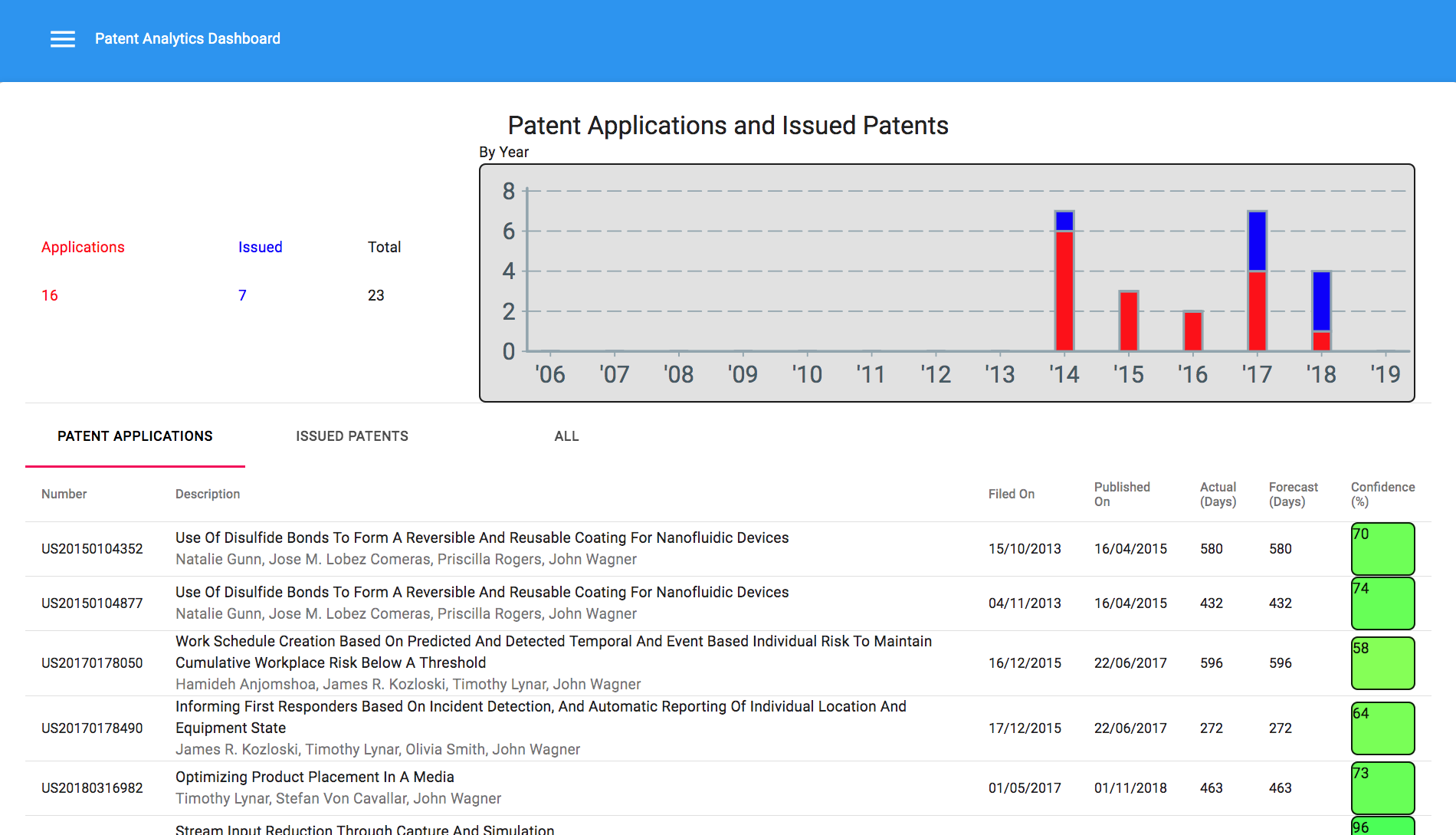}
  \captionof{figure}{Inventor's patent page.}
  \label{fig:inventor}
\end{minipage}%
\begin{minipage}{.5\textwidth}
  \centering
  \includegraphics[width=.95\linewidth]{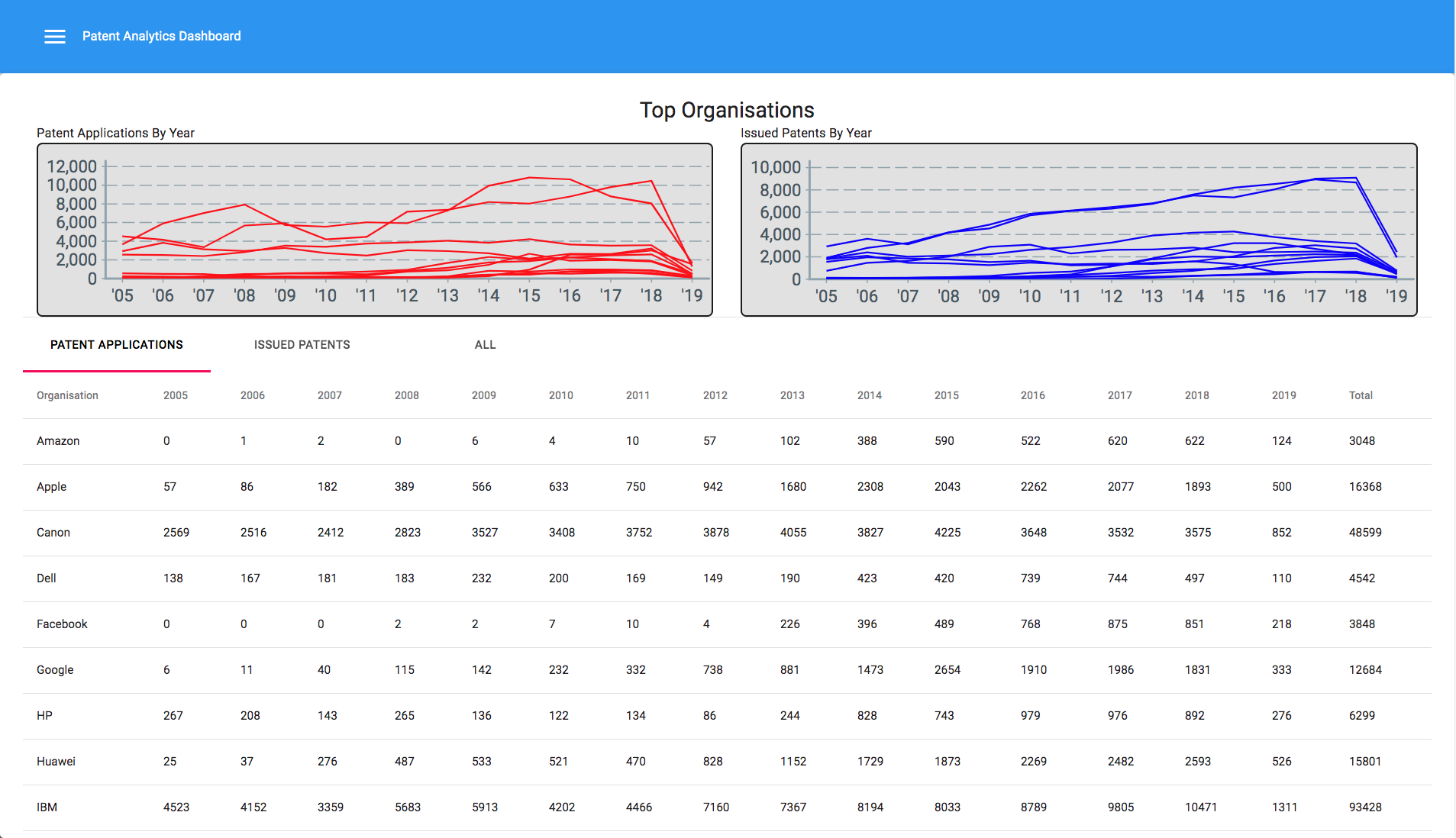}
  \captionof{figure}{Organisation patent summary page.}
  \label{fig:organisations}
\end{minipage}
\end{figure}

\section{Target Users and Demo}
\label{sec:users}

%Three main user groups:
%\begin{itemize}
%  \item Inventors
%  \item Business analysts and managers
%  \item Invention evaluators (internal disclosure teams and at the patent office)
%  \item Patent lawyers
%\end{itemize}

%Current demo support for inventors and business analysts

Our system aims to target inventors, business analysts (and managers), invention evaluators, and patent attorneys. Currently, we have completed the engineering work and UIs catering for inventors and business analysts. We differentiate from current offerings such as Google Patents by focusing on predictive approaches based on machine and deep learning, but can be integrated into existing systems.

Inventors find a page summarising their inventions and applications at the USPTO, as shown in Figure~\ref{fig:inventor}. Our technical contribution here is the underlying machine learning algorithm that predicts, with a prediction confidence score, when a patent application will be granted using features extracted from the patent text. For their granted patents, the user will also find the estimate and the confidence with the variance in prediction coloured from green (good prediction) to red (poor prediction). This allows the user to investigate invention filing time and cause of discrepancies between prediction and results (e.g. a patent maybe worthy of expedited review).

Other pages, as shown in Figure~\ref{fig:organisations} in the demo, present and compare summary statistics of filed patents at targeted organisations. We manually disambiguated over a dozen major technology companies for this work. This view gives an overview of significant inventors within organisations and their output inventions. Similarly, users can perform additional data mining for organisations to find their emerging technology investments and themes of their inventions.

Our future versions showcase underlying deep learning representations of patent text for fast retrieval and comparisons. These representations will allow evaluating patent ideas from short text descriptions (e.g. verifying ideas), clustering patents to find infringing patents and emerging areas (this is currently done via topic modelling and citation graphs), and potentially generating patent titles to seed ideas for inventors.

%\input{sec-related-work.tex}
%\input{sec-dataset.tex}
%\input{sec-features.tex}
%\input{sec-methodology.tex}
%\input{sec-results.tex}
%\input{sec-conclusion.tex}

\bibliographystyle{splncs04}
\bibliography{ecml2019}

\end{document}